\documentclass[floats,floatfix,showpacs,amsmath,amssymb,aps,twocolumn,superscriptaddress,nofootinbib,nolongbibliography,prd]{revtex4-1}

\usepackage{amssymb,amsmath,verbatim,mathtools,needspace,enumitem,etoolbox,graphicx,physics,microtype,afterpage,xspace,tabularx,lmodern,multirow}

\usepackage{gensymb}
\usepackage{soul}
\usepackage{multirow}
\usepackage{subcaption}
\usepackage{ragged2e}
\usepackage{bbm}
\usepackage[dvipsnames, usenames]{xcolor}
\definecolor{linkcolor}{rgb}{0.0,0.3,0.5}
\usepackage[unicode, colorlinks=true, linkcolor=linkcolor, citecolor=linkcolor, filecolor=linkcolor, urlcolor=linkcolor, linktocpage, breaklinks]{hyperref}
\usepackage[all]{hypcap}
\usepackage{cancel}
\usepackage{empheq}
\usepackage[T1]{fontenc}
\usepackage[utf8]{inputenc}
\usepackage[usenames,dvipsnames]{xcolor}
\usepackage{csquotes}
\usepackage{mathrsfs}
\hypersetup{colorlinks=true,citecolor=Periwinkle,linkcolor=Periwinkle,urlcolor=Periwinkle}

\definecolor{romared}{RGB}{142,0,28}

\newcommand{\be}{\begin{equation}}
\newcommand{\ee}{\end{equation}}

\def\be{\begin{equation}}
\def\ee{\end{equation}}
\newcommand{\beq}{\begin{eqnarray}}
\newcommand{\eeq}{\end{eqnarray}}
\usepackage{makecell}
\usepackage{soul}

\newcolumntype{Y}{>{\centering\arraybackslash}X}

\makeatletter
\newcommand*{\rom}[1]{\expandafter\@slowromancap\romannumeral #1@}
\makeatother

\makeatletter
\let\jnl@style=\rm
\def\ref@jnl#1{{\jnl@style#1}}

\def\aj{\ref@jnl{AJ}}                   
\def\actaa{\ref@jnl{Acta Astron.}}      
\def\araa{\ref@jnl{ARA\&A}}             
\def\apj{\ref@jnl{ApJ}}                 
\def\apjl{\ref@jnl{ApJ}}                
\def\apjs{\ref@jnl{ApJS}}               
\def\ao{\ref@jnl{Appl.~Opt.}}           
\def\apss{\ref@jnl{Ap\&SS}}             
\def\aap{\ref@jnl{A\&A}}                
\def\aapr{\ref@jnl{A\&A~Rev.}}          
\def\aaps{\ref@jnl{A\&AS}}              
\def\azh{\ref@jnl{AZh}}                 
\def\baas{\ref@jnl{BAAS}}               
\def\bac{\ref@jnl{Bull. astr. Inst. Czechosl.}}
\def\caa{\ref@jnl{Chinese Astron. Astrophys.}}
\def\cjaa{\ref@jnl{Chinese J. Astron. Astrophys.}}
\def\icarus{\ref@jnl{Icarus}}           
\def\jcap{\ref@jnl{J. Cosmology Astropart. Phys.}}
\def\jrasc{\ref@jnl{JRASC}}             
\def\memras{\ref@jnl{MmRAS}}            
\def\mnras{\ref@jnl{MNRAS}}             
\def\na{\ref@jnl{New A}}                
\def\nar{\ref@jnl{New A Rev.}}          
\def\pra{\ref@jnl{Phys.~Rev.~A}}        
\def\prb{\ref@jnl{Phys.~Rev.~B}}        
\def\prc{\ref@jnl{Phys.~Rev.~C}}        
\def\prd{\ref@jnl{Phys.~Rev.~D}}        
\def\pre{\ref@jnl{Phys.~Rev.~E}}        
\def\prl{\ref@jnl{Phys.~Rev.~Lett.}}    
\def\pasa{\ref@jnl{PASA}}               
\def\pasp{\ref@jnl{PASP}}               
\def\pasj{\ref@jnl{PASJ}}               
\def\rmxaa{\ref@jnl{Rev. Mexicana Astron. Astrofis.}}%
\def\qjras{\ref@jnl{QJRAS}}             
\def\skytel{\ref@jnl{S\&T}}             
\def\solphys{\ref@jnl{Sol.~Phys.}}      
\def\sovast{\ref@jnl{Soviet~Ast.}}      
\def\ssr{\ref@jnl{Space~Sci.~Rev.}}     
\def\zap{\ref@jnl{ZAp}}                 
\def\nat{\ref@jnl{Nature}}              
\def\iaucirc{\ref@jnl{IAU~Circ.}}       
\def\aplett{\ref@jnl{Astrophys.~Lett.}} 
\def\apspr{\ref@jnl{Astrophys.~Space~Phys.~Res.}}
\def\bain{\ref@jnl{Bull.~Astron.~Inst.~Netherlands}} 
\def\fcp{\ref@jnl{Fund.~Cosmic~Phys.}}  
\def\gca{\ref@jnl{Geochim.~Cosmochim.~Acta}}   
\def\grl{\ref@jnl{Geophys.~Res.~Lett.}} 
\def\jcp{\ref@jnl{J.~Chem.~Phys.}}      
\def\jgr{\ref@jnl{J.~Geophys.~Res.}}    
\def\jqsrt{\ref@jnl{J.~Quant.~Spec.~Radiat.~Transf.}}
\def\memsai{\ref@jnl{Mem.~Soc.~Astron.~Italiana}}
\def\nphysa{\ref@jnl{Nucl.~Phys.~A}}   
\def\physrep{\ref@jnl{Phys.~Rep.}}   
\def\physscr{\ref@jnl{Phys.~Scr}}   
\def\planss{\ref@jnl{Planet.~Space~Sci.}}   
\def\procspie{\ref@jnl{Proc.~SPIE}}   

\makeatother

\usepackage{etoolbox} 
\makeatletter
\patchcmd{\@outputpage@head}{\@ifnum{\@mpcol+\@ne}{\@disablepaircolumn}{}}{}{}{}
\makeatother

\begin{document}

\title{Probing Active Galactic Nuclei and Measuring the Hubble constant with Extreme Mass Ratio Inspirals}

\author{Jian-Dong Liu}
\affiliation{School of Fundamental Physics and Mathematical Sciences, Hangzhou Institute for Advanced Study, University of Chinese Academy of Sciences, Hangzhou
310024, China}
\affiliation{Institute of Theoretical Physics, Chinese Academy of Sciences, Beijing 100190, China}
\affiliation{University of Chinese Academy of Sciences,
Beijing 100049, China}

\author{Wen-Biao Han}
\email[]{wbhan@shao.ac.cn}
\affiliation{State Key Laboratory of Radio Astronomy and Technology, Shanghai Astronomical Observatory, Chinese Academy of Sciences, 80 Nandan Road, Shanghai 200030, China}
\affiliation{School of Fundamental Physics and Mathematical Sciences, Hangzhou Institute for Advanced Study, University of Chinese Academy of Sciences, Hangzhou
310024, China}
\affiliation{School of Astronomy and Space Science, University of Chinese Academy of Sciences, Beijing 100049, China}
\affiliation{Taiji Laboratory for Gravitational Wave Universe (Beijing/Hangzhou), University of Chinese Academy of Sciences, Beijing 100049, China}

\author{Hiromichi Tagawa}
\affiliation{Shanghai Astronomical Observatory, Chinese Academy of Sciences, Shanghai 200030, China}

\begin{abstract}
Extreme mass ratio inspirals (EMRIs), consisting of a compact object orbiting a central supermassive black hole, are key targets for future space-based gravitational wave detectors. The millihertz gravitational waves emitted by these systems may carry valuable information about their surrounding astrophysical environments. Over the course of their long-term evolution, interactions between the secondary object and the accretion disk can produce observable effects on both the orbital evolution and the emitted gravitational waveform. Based on the modifications to the companion's orbital evolution induced by the accretion disk environment, we investigate the feasibility of identifying the presence of accretion disk environmental effects in EMRI systems using gravitational wave signals. Within a Bayesian framework, we analyze the capability of EMRI systems with multiple parameter configurations to distinguish accretion disk environmental effects. Our results show that, under the $\alpha$-disk model, all injected events can successfully identify the environment in which the EMRIs reside. Furthermore, we examined the improvement in the precision of Hubble constant measurements using the dark siren method after correctly identifying the accretion disk environment and constraining the relevant disk parameters. Constraining these environmental parameters may further deepen our understanding of the host environment, thereby enabling a more reliable inference of the physical properties of the accretion disk and its associated luminosity and ultimately improving the measurement of cosmological parameters. We find that the measurement precision for a single event can improve by as much as $20\%$. This work highlights the necessity of incorporating environmental effects into future EMRI data analysis. Proper modeling of such effects not only helps identify EMRI systems embedded in accretion disk environments but also further improves the precision of gravitational wave cosmological parameter inference.
\end{abstract}

\maketitle 

\section{Introduction}
Extreme mass ratio inspirals (EMRIs) are binary systems composed of a stellar mass compact object ($\mu \sim 1-100\,M_\odot$) orbiting a supermassive black hole ($M \sim 10^5-10^7\,M_\odot$). The millihertz gravitational waves emitted during their inspiral and orbital precession are among the key targets of future space-based gravitational wave detectors, such as LISA \citep{LISA}, Taiji \citep{Taiji}, and TianQin \citep{Tianqin}. Owing to the extremely small mass ratio $q=\mu/M \ll 1$, an EMRI can undergo approximately $N \sim 1/q$ orbital cycles in the strong-gravity regime, leading to an exceptionally long inspiral phase prior to the final merger and a large number of orbits completed in the vicinity of the supermassive black hole. As a result, EMRIs generate gravitational wave (GW) signals with long durations and extraordinarily rich phase information, making them ideal natural laboratories for precision tests of general relativity \citep{test_gr_1,test_gr_2}, for probing the spacetime properties of supermassive black holes \citep{p_4_BH}, and for inferring cosmological parameters \citep{cosmic_p}.

To date, various formation channels for EMRIs have been proposed. These channels can generally be classified into two categories, depending on whether an accretion disk plays an assisting role in the formation process: the dry channel and the wet channel. In the dry channel, EMRIs primarily form in nuclear star clusters. On the one hand, COs can be scattered into the gravitational wave driven loss cone through multi-body interactions \citep{dry_scattered_1, dry_scattered_2}. On the other hand, via the Hills mechanism \citep{dry_Hill}, a stellar binary can be tidally disrupted in the vicinity of the MBH, with one component gaining energy and being ejected, while the other loses energy and becomes tightly bound to the MBH, eventually evolving into an EMRI. In contrast, the wet channel relies on the presence of an accretion disk. In this scenario, COs initially located outside the disk may interact with the disk, lose orbital energy, and migrate inward \citep{wet_EMRI}. Alternatively, COs may form within the disk and undergo inward migration due to long-term interactions with the surrounding gaseous environment. Although EMRIs formed through the wet channel are expected to exist in only $\sim 1\%$--$10\%$ of galactic nuclei, their formation rate is significantly higher than that of the dry channel \citep{wet_EMRI}. As a result, wet-EMRIs are expected to constitute a substantial fraction of the EMRI sources detectable by LISA \citep{wet_EMRI_1}. Furthermore, in wet-EMRI systems, the presence of an accretion disk implies that EMRIs residing in active galactic nuclei (AGNs) may produce electromagnetic emission coincident with the gravitational wave signal, making them promising candidates for multi-messenger gravitational wave astronomy \citep{multi_EMRI_1, multi_EMRI_2}.

In wet-EMRI systems, a stellar-mass black hole inevitably experiences persistent influences from its surrounding environment throughout its long inspiral \citep{2011_environment_effect_2, 2014_environment_effect_1,2021MNRAS,2023MNRAS}. These include, for example, dynamical friction from a dark-matter halo/spike \citep{DM_effect_1, DM_effect_2}, tidal perturbations induced by nearby objects\citep{tidall_effect}, as well as turbulence, gas drag, and migration torques within an accretion disk \citep{troque_effect_1,troque_effect_2,PRX}. If such environmental effects are neglected in waveform modeling and data analysis, observational results may become entangled with the predictions of vacuum general relativity, introducing systematic biases in parameter estimation and even causing non-negligible contamination in precision tests of strong-field gravity \citep{test_gr_1, test_gr_2, PRX}. Meanwhile, the cumulative impact of environmental effects over long timescales also creates an opportunity for observation: they typically imprint distinctive and potentially detectable features on the gravitational wave signal, most notably a systematic phase drift (dephasing) relative to the vacuum waveform \citep{dephasing_1, dephasing_2}. Conversely, these environmental “fingerprints” embedded in gravitational waves can be used as probes to characterize the astrophysical environment of the source, or instance, to constrain key physical parameters of the accretion disk \citep{PRX, PRD}, reconstruct the mass and density distributions of the nuclear star cluster and/or gas \citep{start_distrbution_1}, and even test whether a dark-matter spike exists near the black hole and measure its profile parameters \citep{DM_effect_1, DM_effect_2}.

In this work, we focus on the final four years before merger in wet EMRI systems and investigate how the accretion environment affects the orbital evolution of the secondary compact object and the resulting gravitational wave signal. We assume that the secondary object is fully embedded in the accretion disk and progrades with respect to the disk gas. We then adopt commonly used power-law prescriptions to characterize the structure of the $\alpha$-disk and incorporate the corresponding disk induced effects as additional terms in the orbital evolution model, thereby constructing a dynamical and waveform description that includes environmental influences. Unlike previous studies on environmental effects in EMRIs, this work further incorporates the impact of eccentricity and orbital inclination evolution on the waveform, allowing us to explore the extent to which the disk scale height and related physical quantities can be constrained. Building on this framework, we perform Bayesian inference for joint parameter estimation over the full parameter space and use model comparison to quantify how well observations can distinguish between dry-EMRI and wet-EMRI scenarios. Given that the disk scale height is closely related to the accretion rate and viscosity parameter of the disk, this opens up new possibilities for using EMRI observations to infer the physical properties of AGN disks and, ultimately, to further constrain cosmological parameters. Finally, under these assumptions, we assess how the inclusion of environmental information enhances the potential of wet EMRIs as dark sirens for cosmological inference and briefly discuss the prospects for identifying electromagnetic counterparts and for wet EMRIs as potential multi-messenger sources.

The structure of this paper is as follows. In Sec.\ref{sec:2}, we introduce the accretion disk model, migration torque model, and EMRI waveform templates adopted in this work, as well as the modifications to the vacuum waveform induced by astrophysical environmental effects. We also present the statistical framework for dark standard sirens with other relevant preliminaries. In Sec.\ref{sec:3}, we present the main results of this paper, including the ability to identify the presence of an accretion disk through environmental effects under different parameter configurations, as well as the impact of adopting different weighting schemes for candidate host galaxies on the precision of the Hubble constant measurement once a wet-EMRI is correctly identified. Finally, Secs.\ref{sec:4} are devoted to a discussion of the main results and a summary. Unless otherwise specified, we adopt geometric units with \(G=c=1\) throughout this paper.

\section{Methods}\label{sec:2}
In this section: Section \ref{sec:2.2.1} introduces the accretion disk model adopted in this work and clarifies how the astrophysical environment affects the orbital evolution of the sBH; Section \ref{sec:2.2.2} presents the waveform templates used and the detailed procedure for Bayesian parameter estimation; Section \ref{sec:2.2.3} introduces the cosmological model adopted in this paper and explains how the Hubble constant is measured using the standard siren approach.


\subsection{Accretion disk}\label{sec:2.2.1}
We assume that a geometrically thin $\alpha$ accretion disk surrounds the SMBH and that the accretion rate onto the SMBH remains approximately constant over the timescales of interest, satisfying \citep{acc_book}

\begin{equation}
    \dot{M} \simeq 3\pi \nu \Sigma 
\end{equation}
where $\nu$ denotes the kinematic viscosity of the disk and $\Sigma$ is the disk surface density. In the $\alpha$-disk framework, $\nu$ is commonly parameterized as

\begin{equation}
    \nu = \alpha_{\rm disk}\, c_s\, H(r)
\end{equation}
where $\alpha_{\rm disk}$ is a dimensionless constant, typically in the range $[0.001,0.1]$ \citep{alpha_range}, $c_s = h\Omega_K$ is the isothermal sound speed, $\Omega_K = M/r^3$ is the Keplerian angular frequency, and $H(r)$ is the disk scale height. In the $\alpha$-disk model, the surface density and aspect ratio of the disk can be written as:

\begin{equation}
\begin{aligned}
\Sigma_{\alpha} ={}& 5.4 \times 10^{2}
\left(\frac{0.1}{\alpha}\right)
\left(\frac{0.1}{f_{\rm Edd}}\frac{\epsilon}{0.1}\right)
\left(\frac{r}{10M}\right)^{3/2} \\
& \times (1 - \sqrt{r_{in}/r})^{-1}
\left[\frac{\mathrm{g}}{\mathrm{cm}^{2}}\right]
\end{aligned}
\label{eq:3}
\end{equation}

\begin{equation}
h_{\alpha} = 0.15
\left(\frac{f_{\rm Edd}}{0.1}\frac{0.1}{\epsilon}\right)
\left(\frac{r}{10M}\right)^{-1}
(1 - \sqrt{r_{in}/r})
\label{eq:4}
\end{equation}

The structural parameters of an $\alpha$ disk can be parameterized in the form of power laws. However, since this work focuses on the orbital evolution of the EMRI during the final four years before merger, when the companion mainly moves in the inner region of the accretion disk, it is necessary to retain the correction terms introduced by the inner boundary condition of the disk in the parametrization \citep{acc_book}. The resulting parametrized equations can be written as follows:

\begin{equation}
\Sigma(r) 
=\Sigma_0
\left( \frac{r}{10M} \right)^{-\Sigma_p} (1 - \sqrt{r_{in}/r})^{-1}\left[\frac{\mathrm{g}}{\mathrm{cm}^{2}}\right]
\label{eq:Sigma_r_wavy}
\end{equation}

\begin{equation}
h(r) 
=h_0 \left( \frac{r}{10M} \right)^{(2\Sigma_p - 1)/4} (1 - \sqrt{r_{in}/r})
\label{eq:h_r_wavy}
\end{equation}
Here, $\Sigma_{0}$ and $h_{0}$ are chosen as representative inner disk values inferred from astrophysical observations, with $\Sigma_{0}\in[10^{3},\,10^{6}]$ and $h_{0}\in[0.01,\,0.1]$ . For an $\alpha$ disk, we set $\Sigma_{p}=-3/2$. $r_{in}$ denotes the distance between the inner edge of the accretion disk and the central black hole. In this work, we take the radius of the innermost stable circular orbit (ISCO) as the value of $r_{in}$.

Before constructing the torque (or the effective drag) exerted by the accretion disk on the sBH, one should compare the relative velocity of the sBH with respect to the disk gas, \(v_{\rm rel}\), to the local sound speed \(c_s\), since the form of the environmental interaction differs between the subsonic and supersonic regimes. In general, \(v_{\rm rel}\) receives contributions from both the orbital eccentricity and the orbital inclination with respect to the disk. We adopt the approximation:

\begin{equation}
v_{\rm rel} \sim v_{K}\sqrt{e^2+\sin^2\iota_{\rm sd}}
\sim r\Omega_K\sqrt{e^2+\sin^2\iota_{\rm sd}}
\end{equation}
where \(\iota_{\rm sd}\) is the angle between the sBH orbital plane and the disk mid-plane, and \(v_K=r\Omega_K\) is the Keplerian velocity. We assume that once the EMRI enters the LISA band (corresponding to the inner disk region), the sBH is fully embedded in the accretion disk and satisfies \(\iota_{\rm sd}<h\), with \(h\equiv H/r\) the disk aspect ratio. Moreover, since in wet-EMRIs the inclination damps rapidly and the orbit circularizes efficiently, both \(\iota_{\rm sd}\) and \(e\) are typically small \citep{wet_EMRI, low_e}. The Mach number is then

\begin{equation}
\mathcal{M}\equiv \frac{v_{\rm rel}}{c_s}\simeq
\frac{r\Omega_K\sqrt{e^2+\sin^2\iota_{\rm sd}}}{h\,r\Omega_K}
=\frac{\sqrt{e^2+\sin^2\iota_{\rm sd}}}{h}
\end{equation}

Under typical conditions, the migration regime of the sBH is determined by the Mach number \(\mathcal{M}\): when \(\mathcal{M}>1\), the migration occurs in the supersonic regime, whereas when \(\mathcal{M}<1\), it occurs in the subsonic regime. Since the value of \(\mathcal{M}\) depends on the specific parameter configuration, we adopt the prescription proposed by \cite{2020MNRAS} to evaluate the impact of the companion's migration in the accretion disk on the evolution of its orbital parameters. This prescription is based on dynamical friction theory and is applicable to both supersonic and subsonic regimes. In the small inclination case (\(\iota_{\rm sd}<h\)),the environmental impact on the orbital evolution can be characterized by the following inverse timescales:

\begin{equation}
\tau_e^{-1} = -\frac{de/dt}{e} = 0.780\,t_{\rm wave}^{-1}
\left[1+\frac{1}{15}\left((\frac{e}{h})^{2}+ (\frac{\iota_{sd}}{h})^{2}\right)^{3/2}\right]^{-1}
\end{equation}

\begin{equation}
\tau_i^{-1} = - \frac{d\iota_{sd}/dt}{\iota_{sd}} = 0.544\,t_{\rm wave}^{-1}
\left[1+\frac{1}{21.5}\left((\frac{e}{h})^{2}+(\frac{\iota_{sd}}{h})^{2}\right)^{3/2}\right]^{-1}
\end{equation}

\begin{equation}
\tau_a^{-1} = - \frac{da/dt}{a} = 4.25 h^{2} t_{\rm wave}^{-1}
\left[1+\frac{4.25}{21}\left((\frac{e}{h})^{2}+(\frac{\iota_{sd}}{h})^{2}\right)^{1/2}\right]^{-1}
\end{equation}
where \(t_{\rm wave}^{-1}\) is the inverse characteristic migration timescale, given by

\begin{equation}
t_{\rm wave}^{-1}=\frac{\mu}{M}\left(\frac{\Sigma r^{2}}{M}\right)h^{-4}\Omega_K
\end{equation}

\subsection{EMRI Waveforms}\label{sec:2.2.2}
In the radiation zone far from the gravitational-wave source, the gravitational wave signal can be written in a complex, dimensionless time domain strain, where \(h_+\) and \(h_\times\) denote the two polarization components in the transverse--traceless (TT) gauge \citep{few_waveform}:

\begin{equation}
h \equiv h_{+}- i h_{\times}
= \frac{\mu}{d_L}\sum_{l m k n}
A_{l m k n}(t)\, S_{l m k n}(t,\theta)\,
e^{i m \phi}\, e^{-i\Phi_{mkn}(t)} 
\end{equation}
Here \(d_L\) is the luminosity distance of the source, and \(t\) is the time at which the gravitational wave arrives at the Solar System barycenter (SSB). The angles \(\theta\) and \(\phi\) are the polar and azimuthal viewing angles in the source frame. For a given mode \((m,k,n)\), the phase can be decomposed as $\Phi_{mkn}(t)=m\,\Phi_{\phi}(t)+k\,\Phi_{\theta}(t)+n\,\Phi_{r}(t)$ i.e., an integer linear combination of the azimuthal, polar, and radial phases. The mode amplitude \(A_{l m k n}\) is related to the Teukolsky mode amplitude \(Z_{l m k n}\) in the asymptotic region by $A_{l m k n}(t)=-\frac{2\,Z_{l m k n}(t)}{\omega_{mkn}^2(t)} , $ where $\omega_{mkn}(t)=m\,\Omega_{\phi}(t)+k\,\Omega_{\theta}(t)+n\,\Omega_{r}(t)$ is the angular frequency of that mode, and \(\Omega_r\), \(\Omega_{\theta}\), and \(\Omega_{\phi}\) are the fundamental radial, polar, and azimuthal frequencies of bound Kerr geodesic orbits, respectively. The function \(S_{l m k n}(t,\theta)\) denotes the corresponding spin-weighted spheroidal harmonic. The indices \(l,m,k,n\) label the orbital angular momentum mode, the azimuthal mode, the polar, and radial modes, respectively.

During the evolution of an EMRI, the radiation-reaction timescale is typically much longer than the orbital period timescale, so an adiabatic approximation can be adopted: the inspiral is treated as a slow drift through a sequence of instantaneous Kerr geodesic orbits, which provides a convenient description of the gravitational wave emission. For bound orbits, the orbital shape and orientation are uniquely characterized by three constants of motion (per unit mass), $\mathbf I \equiv (E,\;L_z,\;Q)$, so specifying the rates of change of \(\mathbf I\) is sufficient to determine the time evolution of the orbital parameters. Because the environment surrounding an EMRI (e.g., an accretion disk) may interact with the sBH over long times and exert additional non-conservative forces, we write the evolution of the constants of motion as the sum of the gravitational wave radiation-reaction contribution and an environmental contribution:

\begin{equation}
\frac{d\mathbf I}{dt}
= \left(\frac{d\mathbf I}{dt}\right)_{\rm GW} + \,\left(\frac{d\mathbf I}{dt}\right)_{\rm env}
\end{equation}

To describe more transparently how environmental effects impact the orbital geometry, we further map the evolution of the constants of motion to Keplerian-like orbital parameters, $\mathbf x\equiv(p,\;e,\;\iota)$. Since \(\mathbf I=\mathbf I(\mathbf x)\), taking a time derivative yields:

\begin{equation}
\dot{\mathbf x}
= \mathbf J^{-1}(\mathbf x)\left(\frac{d\mathbf I}{dt}\right)_{\rm GW} +
\,\mathbf J^{-1}(\mathbf x)\left(\frac{d\mathbf I}{dt}\right)_{\rm env}
\end{equation}
where \(\mathbf J(\mathbf x)\) is the Jacobian matrix,

\begin{equation}
\mathbf J(\mathbf x)=\frac{\partial(E,\;L_z,\;Q)}{\partial(p,\;e,\;\iota)}
\end{equation}

These relations allow us, once prescriptions for \(\left(d\mathbf I/dt\right)_{\rm GW}\) and \(\left(d\mathbf I/dt\right)_{\rm env}\) are specified, to directly obtain the adiabatic evolution of \((p,e,\iota)\), thereby propagating environmental effects consistently into the orbital fundamental frequencies and the waveform phase. Environmental effects can alter the orbital evolution of an EMRI system, thereby inducing a cumulative phase shift in the gravitational wave signal.  Therefore  the dephasing is:

\begin{equation}
\Delta \phi \equiv \phi_{\mathrm{vacuum}} - \phi_{\mathrm{non\text{-}vacuum}} 
\end{equation}
Here, $\Phi_{\mathrm{vacuum}}$ and $\Phi_{\mathrm{non\text{-}vacuum}}$ denote the gravitational wave phases calculated using the vacuum model and the model including environmental effects, respectively.

In this work, we perform parameter inference within a Bayesian framework. According to Bayes theorem, the posterior distribution is given by:

\begin{equation}
    p(\theta | d) = \frac{\mathcal{L}(d|\theta)\pi(\theta)}{\int d\theta \mathcal{L}(d|\theta)\pi(\theta)}=
    \frac{\mathcal{L}(d|\theta)\pi(\theta)}{\mathcal{Z}}
\end{equation}
where $d$ denotes the detector strain data, $\theta$ is the set of signal parameters to be inferred, and $h(\theta)$ is the model waveform evaluated at $\theta$. $\pi(\theta)$ is the proir distribution for $\theta$; see Table \ref{tab:priors}. $\mathcal{Z}$ is a normalization factor which called "evidence", and given the detector noise power spectral density (PSD) $S_n(f)$, we construct the likelihood function $\mathcal{L}$ for the gravitational-wave data. The likelihood is written as

\begin{equation}
    \ \mathcal{L}(d|\theta)
= exp[-\frac{1}{2}\,(d-h(\theta) \,|\, d-h(\theta)) \,]
\end{equation}

Furthermore, to quantify whether the formation channel of an EMRI can be identified through the environmental effects around it, we employ the Bayes factor to compare the \textit{vacuum} and \textit{non-vacuum} scenarios, defined as

\begin{equation}
    \ln BF_{\mathrm{vacuum}}^{\mathrm{non\text{-}vacuum}}
= \ln\left(\mathcal{Z}_{\mathrm{non\text{-}vacuum}}\right)-\ln\left(\mathcal{Z}_{\mathrm{vacuum}}\right)
\end{equation}

When \(\left|\ln BF_{\mathrm{vacuum}}^{\mathrm{non\text{-}vacuum}}\right| < 2.4\), the current evidence is considered insufficient to effectively distinguish whether the EMRI event is affected by environmental effects. When \(\ln BF_{\mathrm{vacuum}}^{\mathrm{non\text{-}vacuum}} > 2.4\), the observational results favor the interpretation that the EMRI event is influenced by environmental effects; otherwise, vice versa.

We adopt the noise-weighted inner product defined as:

\begin{equation}
    (a|b)=4\,\mathrm{Re}\int_{0}^{\infty}\frac{\tilde a(f)\,\tilde b^{*}(f)}{S_n(f)}\,\mathrm{d}f \,
\end{equation}
where $\tilde a(f)$ is the Fourier transform of the time series $a(t)$, $^{*}$ denotes complex conjugation, and $S_n(f)$ is the detector noise power spectral density.

The signal-to-noise ratio (SNR) of a gravitational wave signal is defined by

\begin{equation}
    \rho=\sqrt{(h|h)}\,
\end{equation}

In this work, we classify a signal as detectable when $\rho>15$.

Table \ref{tab:priors} summarizes the parameter definitions and prior choices used in this analysis. We then inject these signals into Gaussian noise generated according to the LISA design sensitivity and use the Eryn sampler to infer the posterior distributions of the model parameters \citep{eryn}.

\begin{table*}[htbp]
\centering
\caption{Parameters and priors.}
\label{tab:priors}
\begin{tabular}{lll}
\hline
Parameter & Definition & Prior \\
\hline
$M$ & SMBH mass ($M_\odot$) & Log-uniform in $[10^{5},\,10^{7}]\,M_\odot$ \\
$\mu$ & sBH mass ($M_\odot$) & Uniform in $[10,\,100]\,M_\odot$ \\
$a$ & SMBH dimensionless spin magnitude & Uniform in $[0,\,0.99]$ \\
$p_{0}$ & Initial semi-latus rectum in units of $M$ ($p/M$) & Uniform in $[6,\,20]$ \\
$e_{0}$ & Initial orbital eccentricity & Uniform in $[0,\,0.2]$ \\
$\cos\iota_{0}$ & Initial inclination cosine & Uniform in $[-1,\,1]$ \\
$D_{L}$ & Luminosity distance (Gpc) & Uniform in $[0.1,\,10]$~Gpc \\
$\theta_{S}$ & Sky-location polar angle & Uniform in $[0,\,\pi]$ \\
$\phi_{S}$ & Sky-location azimuthal angle & Uniform in $[0,\,2\pi)$ \\
$\theta_{K}$ & Polar angle of the MBH spin direction & Fixed to $1.2$ \\
$\phi_{K}$ & Azimuthal angle of the MBH spin direction & Fixed to $0.6$ \\
$\Phi_{\phi,0}$ & Initial azimuthal phase & Fixed to $2$ \\
$\Phi_{\theta,0}$ & Initial polar phase & Fixed to $3$ \\
$\Phi_{r,0}$ & Initial radial phase & Fixed to $4$ \\
$\Sigma_{0}$ & Surface-density normalization factor & Log-uniform in $[10^{3},\,10^{6}]\,\mathrm{g\,cm}^{-2}$ \\
$h_{0}$ & Aspect-ratio normalization factor & Uniform in $[10^{-2},\,10^{-1}]$ \\
\hline
\end{tabular}
\vspace{1mm}
\begin{minipage}{0.95\textwidth}
\small
\textit{Notes.} The injected value of $p_{0}$ is determined from the orbital evolution during the final four years before merger; the injected value of the luminosity distance $D_L$ is set by tuning the SNR to represent both low- and high-SNR cases; all other injected parameters are randomly chosen. Angles are given in radians. ``Log-uniform'' denotes a prior flat in $\log_{10}$ of the parameter.
\end{minipage}
\end{table*}

\subsection{Cosmological Models and Standard Sirens}\label{sec:2.2.3}
\subsubsection{$\Lambda$CDM Cosmological Model}
In this study, we adopt the flat $\Lambda$CDM model as our baseline cosmology, in which the expansion of the Universe is characterized by the Hubble--Lema\^{i}tre parameter $H(z)$:

\begin{equation}
H(z)=H_0\sqrt{\Omega_m(1+z)^3+(1-\Omega_m)}\,
\end{equation}
where the dark energy equation-of-state (EoS) parameter is assumed to be $w=-1$, and $\Omega_m$ is the present day matter density parameter \citep{DM_density}. The fiducial values of the remaining cosmological parameters used in our simulations are based on the \textit{Planck} 2018 results (TT, TE, EE + lowE), adopting $\Omega_m=0.3166$ and $H_0 = 67.27^{+0.60}_{-0.60}\ {\rm km\ s^{-1}\ Mpc^{-1}}\,$ \citep{Planck2018}.

In addition, the luminosity distance of a source at redshift $z$ is given by

\begin{equation}
d_L(z)=(1+z)c\int_{0}^{z}\frac{dz'}{H(z')}\,
\label{dl}
\end{equation}
where $c$ is the speed of light and $H(z)$ is the Hubble parameter describing the expansion rate of the Universe at redshift $z$.

\subsubsection{Dark Sirens Statistical Framework}
The key to inferring cosmological parameters using the dark-siren method is to statistically associate the three-dimensional localization volume measured from GW events with galaxy catalogs obtained by electromagnetic surveys. In this work, we adopt the hierarchical Bayesian inference frameworks proposed by \cite{Standard_sirens_1} and \cite{Standard_sirens_2}. For a set of $N$ mutually independent GW-event datasets $D_{\rm GW}\equiv\{d_{{\rm GW},i}\}_{i=1}^{N}$ and the corresponding EM datasets $D_{\rm EM}\equiv\{d_{{\rm EM},i}\}_{i=1}^{N}$ the posterior distribution of the Hubble constant $H_0$ can be written as

\begin{equation}
p(H_0\mid D_{\rm GW},D_{\rm EM})
\propto p_0(H_0)\prod_{i=1}^{N}p\!\left(d_{{\rm GW},i},d_{{\rm EM},i}\mid H_0\right)
\end{equation}
where $p_0(H_0)$ is the prior on $H_0$. In this work, we adopt a uniform prior $H_0\in[20,140]~{\rm km\,s^{-1}\,Mpc^{-1}}$. Here $p(d_{{\rm GW},i},d_{{\rm EM},i}\mid H_0)$ denotes the joint GW--EM likelihood for the $i$-th event. Because the GW and EM data are conditionally independent given the source parameters, the joint likelihood can be factorized into a GW likelihood term and an EM likelihood term, followed by marginalization over nuisance parameters, including the sky position $\Omega\equiv(\alpha,\delta)$, luminosity distance $D_L$, redshift $z$, and luminosity $L$. This yields

\begin{equation}
\begin{aligned}
p(d_{{\rm GW},i},d_{{\rm EM},i}\mid H_0)
&= \frac{1}{\beta(H_0)}
\int p(d_{{\rm GW},i},d_{{\rm EM},i},\Omega,D_L,z,L\mid H_0)
\\
&\quad \times {\rm d}\Omega\,{\rm d}D_L\,{\rm d}z\,{\rm d}L \,.
\end{aligned}
\end{equation}
where $\beta(H_0)$ is a normalization factor accounting for GW and EM selection effects.

The integrand can be further decomposed as

\begin{equation}
\begin{aligned}
p(&d_{{\rm GW},i},d_{{\rm EM},i},\Omega,D_L,z,L \mid H_0)
\\
&= p(d_{{\rm GW},i},d_{{\rm EM},i}\mid \Omega,D_L,z,L,H_0)\,
   p_0(\Omega,D_L,z,L\mid H_0)
\\
&= p(d_{{\rm GW},i}\mid \Omega,D_L)\,
   p(d_{{\rm EM},i}\mid \Omega,z,L)
\\
&\quad \times \delta\!\bigl(D_L-\hat D_L(z,\Omega)\bigr)\,
   p_0(z,\Omega,L\mid H_0)
\\
&= p(d_{{\rm GW},i}\mid \Omega,\hat D_L(z,\Omega))\,
   p(d_{{\rm EM},i}\mid \Omega,z,L)\,
   p_0(z,\Omega,L\mid H_0)\,.
\end{aligned}
\end{equation}
where we have used $\delta\!\big(D_L-\hat D_L(z,\Omega)\big)$. Here $\hat D_L(z,\Omega)$ denotes the luminosity distance at redshift $z$ (computed using Eq.~\eqref{dl}) under the assumed cosmology.%

Under a Gaussian approximation, the marginalized GW likelihood for a single event is written as

\begin{equation}
p\!\left(d_{{\rm GW},i}\mid \hat D_L(z,\Omega)\right)
\propto
\exp\!\left[-\frac{1}{2}\frac{\left(D_L-\hat D_L(z,\Omega)\right)^2}{\sigma_{D_L}^2}\right]
\end{equation}
The corresponding EM likelihood can be expressed as a weighted sum over candidate host galaxies within the localization volume,

\begin{equation}
p\!\left(d_{{\rm EM},i}\mid \Omega,z,L\right)
\propto
\frac{1}{N_{\rm gal}}
\sum_{j=1}^{N_{\rm gal}}
W_j\,
\exp\!\left[-\frac{1}{2}\frac{(z_j-z_{\rm gw})^2}{\sigma_{z_j}^2}\right]
\end{equation}
where $N_{\rm gal}$ is the number of candidate host galaxies; $z_{\rm gw}$ denotes the redshift inferred from the GW luminosity-distance information for a given $H_0$; and $\sigma_{z_j}$ is the redshift uncertainty of the $j$-th candidate host galaxy (as provided by the catalog).

\begin{equation}
\sigma_{z_j}=\sqrt{\left(\sigma_{z}^{\rm cat}\right)^2+\left(\sigma_{z}^{\rm PV}\right)^2}\,
\end{equation}
where $\sigma_{z}^{\rm cat}$ is the catalog redshift measurement uncertainty, and the redshift uncertainty induced by the host peculiar velocity is taken to be

\begin{equation}
\sigma_{z}^{\rm PV}(z)=(1+z)\frac{\sqrt{\langle v^2\rangle}}{c}\,
\end{equation}
with $\sqrt{\langle v^2\rangle}$ characterizing the galaxy velocity dispersion (velocity uncertainty). In this work, we adopt $\sqrt{\langle v^2\rangle}=500\,{\rm km\,s^{-1}}$. In addition, we introduce a weight $W_j$ to quantify the consistency of each candidate host in terms of sky position and luminosity information, and we adopt:

\begin{equation}
W_j \propto W^{\rm pos}_j\,W^{\rm lum}_j
=\exp\!\left[-\frac{1}{2}\left(\chi^2_{{\rm pos},j}+\chi^2_{{\rm lum},j}\right)\right]
\end{equation}
The positional consistency term is defined using the two-dimensional covariance of the GW sky localization. Let the GW source sky position be $(\bar\theta,\bar\phi)$ and the angular position of the $j$-th candidate host be $(\theta_j,\phi_j)$. We define the offset vector $\xi_j=(\theta_j-\bar\theta,\;\phi_j-\bar\phi)$ and extract the corresponding $2\times2$ sub-matrix from the full GW covariance as $\mathrm{Cov}_{\rm sky}\equiv\mathrm{Cov}[\theta,\phi]$. The Mahalanobis distance is then $\chi^2_{{\rm pos},j} =\xi_j^{\mathsf T}\, \mathrm{Cov}_{\rm sky}^{-1}\,\xi_j$.

The luminosity consistency term quantifies the agreement between the observed luminosity in the catalog and the luminosity predicted by our model. For the $j$-th candidate host, the catalog provides an observed bolometric luminosity $L^{\rm cat}_j$, while our method yields a calculated bolometric luminosity $L^{\rm calc}_j = f^{j}_{Edd}L^{j}_{Edd} = f^{j}_{Edd} 1.26\times10^{38}(M/M_{\odot})erg/s$. We define the residual in log-luminosity space as

\begin{equation}
\Delta \ell_j \equiv \log_{10}L^{\rm cat}_j-\log_{10}L^{\rm calc}_j
\end{equation}
and assume $\Delta \ell_j$ follows a Gaussian distribution with zero mean, so that

\begin{equation}
\chi^2_{{\rm lum},j}=\frac{\Delta \ell_j^2}{\sigma_{\ell,j}^2},
\qquad
\sigma_{\ell,j}^2=\sigma_{\ell,{\rm cat},j}^2+\sigma_{\ell,0}^2
\end{equation}
where $\sigma_{\ell,{\rm cat},j}$ is the observational uncertainty from the catalog and $\sigma_{\ell,0}$ is an additional bolometric-scatter term.

In the above decomposition, $p_0(z,\Omega,L\mid H_0)$ denotes the prior distribution of host galaxies in the $(z,\Omega,L)$ space. We assume galaxies are uniformly distributed in comoving volume. Because the dark-siren inference relies on galaxy catalogs, catalog incompleteness can modify the effective redshift distribution of host galaxies and thus affect the constraint on $H_0$. To account for this effect, we model the host prior as a weighted mixture of ``cataloged'' and ``missing'' populations,

\begin{equation}
p_0(z,\Omega\mid H_0)=
f\,p_{\rm cat}(z,\Omega \mid H_0)
+(1-f)\,p_{\rm miss}(z,\Omega\mid H_0)
\label{eq:p0_mixture}
\end{equation}

Here, $p_{\rm cat}$ denotes the probability distribution of catalogued sources in the $(z,\Omega)$ space, while
$p_{\rm miss}$ denotes the probability distribution of sources missing from the catalog in the $(z,\Omega)$ space.
They can be written as:

\begin{equation}
p_{\mathrm{cat}}(z,\Omega)
=
\frac{n_{\mathrm{cat}}(z,\Omega)}{N_{\mathrm{cat}}(R)}
\frac{dV_c}{dz\,d\Omega}\,
\end{equation}

\begin{equation}
p_{\mathrm{miss}}(z,\Omega)
=
\frac{n_{\mathrm{miss}}(z,\Omega)}{N_{\mathrm{miss}}(R)}
\frac{dV_c}{dz\,d\Omega}\,
\end{equation}

Here $N_{\rm cat}(R)$ is the total number of sources within the localization volume $R$ that are included in the catalog,
and $N_{\rm miss}(R)$ is the total number of sources within $R$ that are not included in the catalog, satisfying

\begin{equation}
N_{\rm miss}(R)=N_{\rm tot}(R)-N_{\rm cat}(R)
\end{equation}

where $N_{\rm tot}(R)$ denotes the total number of sources that are expected to exist in the localization region $R$ in principle. Correspondingly, $n_{\rm cat}(z,\hat\Omega)$ is the comoving number density of catalogued sources, and $n_{\rm miss}(z,\hat\Omega)$ is the comoving number density of sources missing from the catalog. In practice, we compute

\begin{equation}
n_{\rm miss}(z,\Omega)=n_{\rm tot}(z,\Omega)-n_{\rm cat}(z,\Omega)
\end{equation}

where $n_{\rm tot}(z,\hat\Omega)$ is the underlying comoving number density of AGN sources. Based on existing observational results \citep{AGN_density}, we adopt

\begin{equation}
n_{\rm tot}=2\times 10^{-4}\ {\rm Mpc}^{-3}
\end{equation}

Accordingly, the completeness fraction can be approximated as
\begin{equation}
f = \frac{1}{V_{c}(z_{max})} \int p_{cat} \frac{dV_{c}}{dzd\Omega}dzd\Omega
\end{equation}

Finally, to correct for selection effects introduced by the instrumental sensitivity and detection pipelines for both GW events and galaxy samples, we follow \cite{Standard_sirens_1}and introduce the normalization factor $\beta(H_0)$ in the hierarchical Bayesian inference. This factor is determined by the GW detection condition and the prior and can be written as

\begin{equation}
\begin{aligned}
\beta(H_0)
&= \int_{\rho>\rho_{\rm th}}
{\rm d}d_{\rm GW}\,{\rm d}\Omega\,{\rm d}z\,{\rm d}L
\\
&\quad \times p\!\left(d_{\rm GW}\mid \hat D_L(z,\Omega)\right)\,
p_0(z,\Omega,L\mid H_0) \,.
\end{aligned}
\end{equation}
where $\rho_{\rm th}$ is the SNR threshold.

\section{Results}\label{sec:3}
In our analysis, we adopted the $\alpha$-disk model as the benchmark accretion-disk model, mainly because it is one of the most commonly used standard models in studies of AGN accretion disks. Within this framework, the surface density, aspect ratio, accretion rate, and viscosity parameters of the disk are linked through relatively clear physical relations, which makes it convenient to translate the environmental constraints obtained from GW observations into constraints on the physical properties of the accretion disk and its potential electromagnetic luminosity. In addition, the relatively small number of free parameters in this model is advantageous for controlling parameter degeneracy in the Bayesian analysis, allowing us to assess more clearly the impact of environmental effects on EMRI signals and cosmological parameter inference.

In this section, we present a systematic analysis and discussion of the relevant results. We begin by illustrating the impact of environmental effects of different strengths on EMRI waveforms. We then analyze injected EMRI events embedded in accretion disks using both vacuum and environmental templates and compare the corresponding parameter estimation results in the two cases. Their ability to distinguish between the two scenarios is further quantified using the Bayes factor. In addition, we investigate the extent to which the environmentally corrected templates can constrain the corresponding accretion disk parameters under different disk conditions. On this basis, within the dark siren framework, we assess the improvement in the precision of the Hubble constant inferred from a single event when the measured accretion disk parameters are used to estimate the luminosities of candidate host galaxies and assign them as weights, compared with the case where only sky location weighting is considered.

\subsection{Capability of distinguishing the surrounding astrophysical environment}
\begin{figure*}[ht!]
\centering
\includegraphics[width=0.9\linewidth]{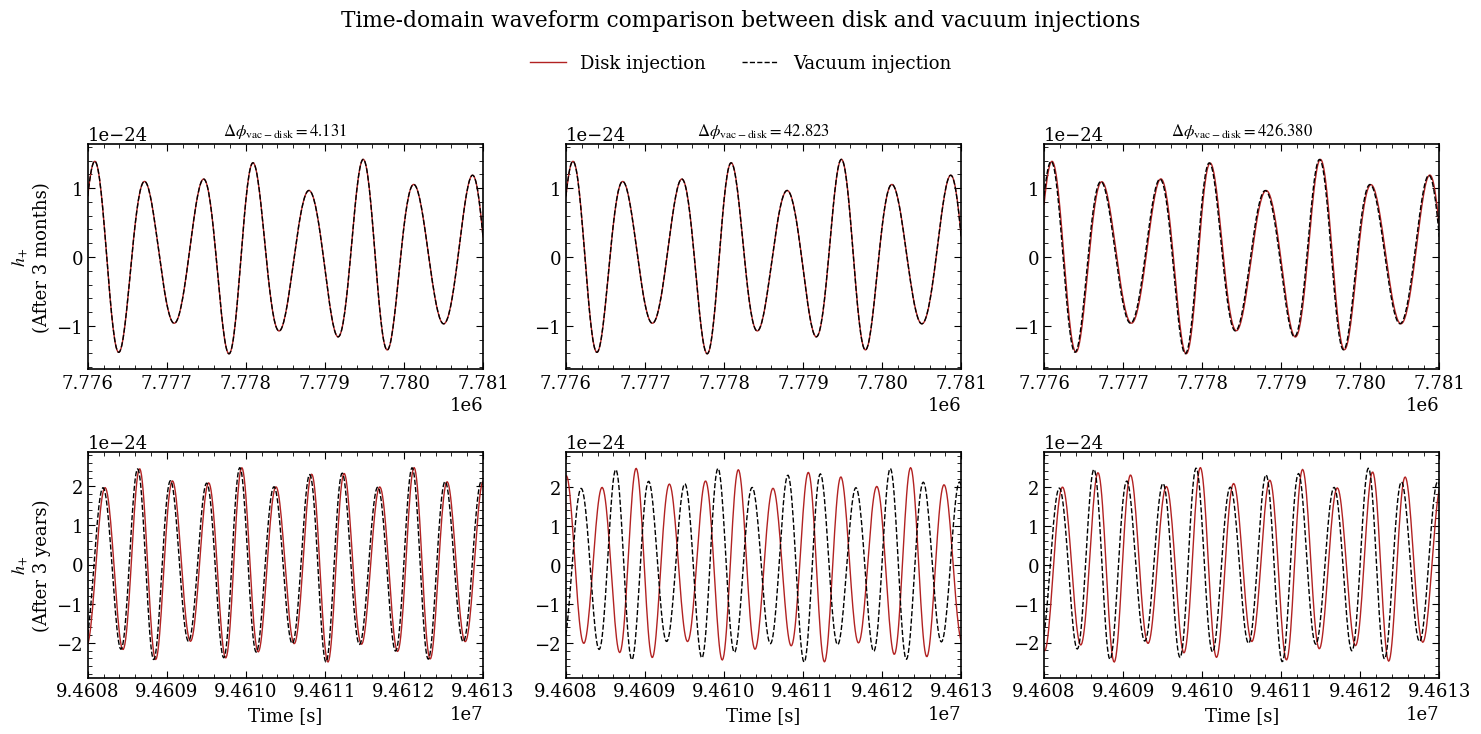}
\caption{Time-domain comparison of EMRI waveforms with environmental corrections and in the vacuum case. The red solid lines denote the waveforms, including environmental effect corrections, whereas the black dashed lines represent the vacuum waveforms. From left to right, the three columns correspond to $\Delta\phi_{\rm vac-disk}=4.131$, 42.823, and 426.38, respectively. The top and bottom rows show the local waveform segments after 3 months and 3 years of evolution, respectively.
\label{fig:1}}
\end{figure*}

Figure \ref{fig:1} provides an intuitive illustration of how the environmental effects of different strengths influence the phase evolution of EMRI waveforms. We inject an EMRI event with parameters $M=10^6$, $\mu=20$, and $e=0.05$, and by adjusting $\Sigma_0$ and $h_0$, we obtain three cases with dephasings of 4.131, 42.823, and 426.38 from left to right, respectively. The figure compares the waveform phases when the system first enters the LISA sensitivity band, after evolving for 3 months, and after evolving for 3 years. It can be seen that, during the early inspiral stage, phase perturbations induced by environmental effects of different strengths remain insignificant. However, as the small compact object continues to inspiral under the influence of the disk environment over a long timescale, these initially tiny differences accumulate and are amplified, eventually leading to clearly distinguishable dephasing features in the late stage evolution.

\begin{figure*}[ht!]
\centering
\includegraphics[width=0.9\linewidth]{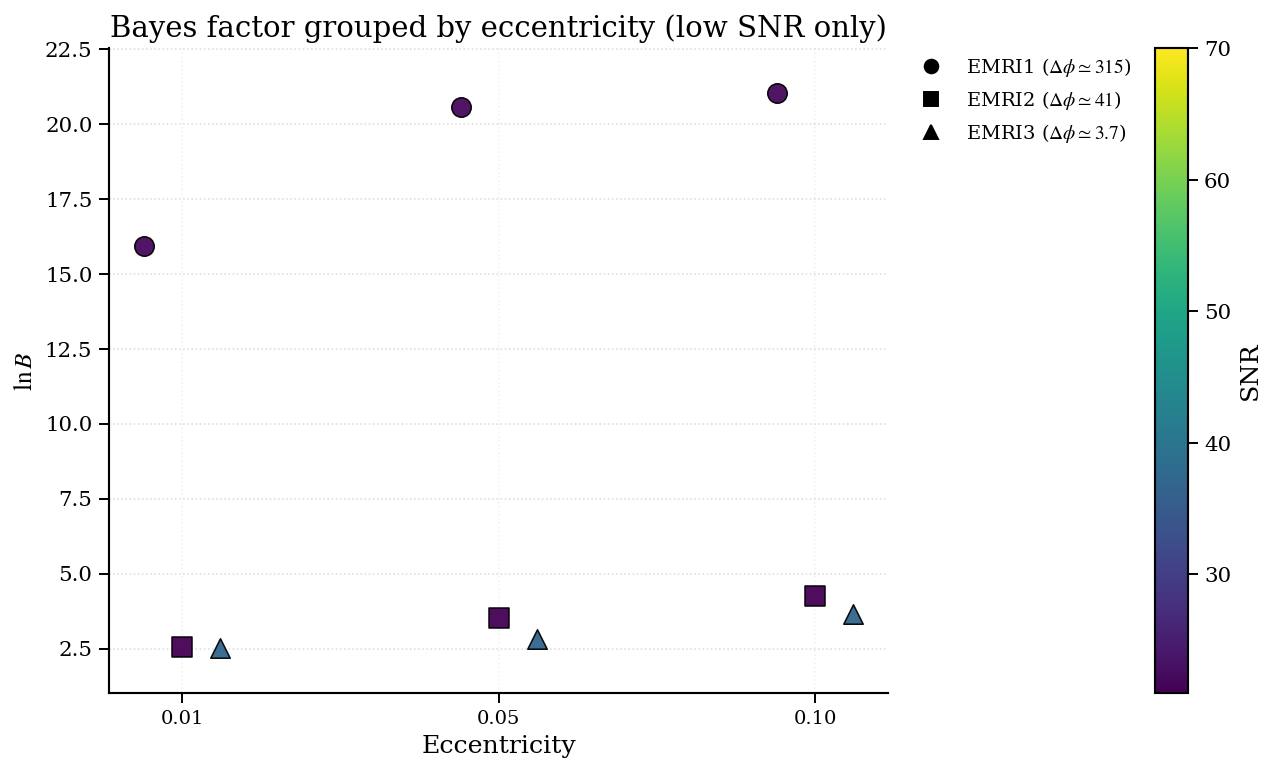}
\caption{The figure shows the logarithmic Bayes factor, $\ln B$, for different EMRI systems under low-SNR and high-SNR conditions with eccentricity $e = 0.10$. Different marker shapes correspond to three EMRI signals: circles denote EMRI1 ($\Delta\phi \approx 315$ radias), squares denote EMRI2 ($\Delta\phi \approx 41$ radians), and triangles denote EMRI3 ($\Delta\phi \approx 3.7$ radians). The colors indicate the corresponding SNR values. 
\label{fig:2}}
\end{figure*}

Figure~\ref{fig:2} illustrates LISA's capability, over a 4-year observation period, to distinguish whether an EMRI is embedded in an accretion disk environment under different SNR conditions. We consider three representative EMRI events, for which the cumulative dephasing induced by environmental effects is 315, 41, and 3.7, respectively. The results show that, for all three events, the Bayes factors indicate that the presence or absence of an accretion disk environment can be effectively identified in both low and high SNR cases. Furthermore, as the SNR increases, the Bayes factor generally becomes larger, indicating stronger support from model selection for the presence of an accretion disk environment. This suggests that a higher SNR enables a more effective extraction of waveform features associated with environmental effects, making the long term phase accumulation and subtle waveform corrections induced by the accretion disk environment easier to resolve. Therefore, at higher SNR, the model including accretion disk environmental effects can obtain stronger support than the vacuum model.

\begin{figure*}[ht!]
\centering
\includegraphics[width=0.9\linewidth]{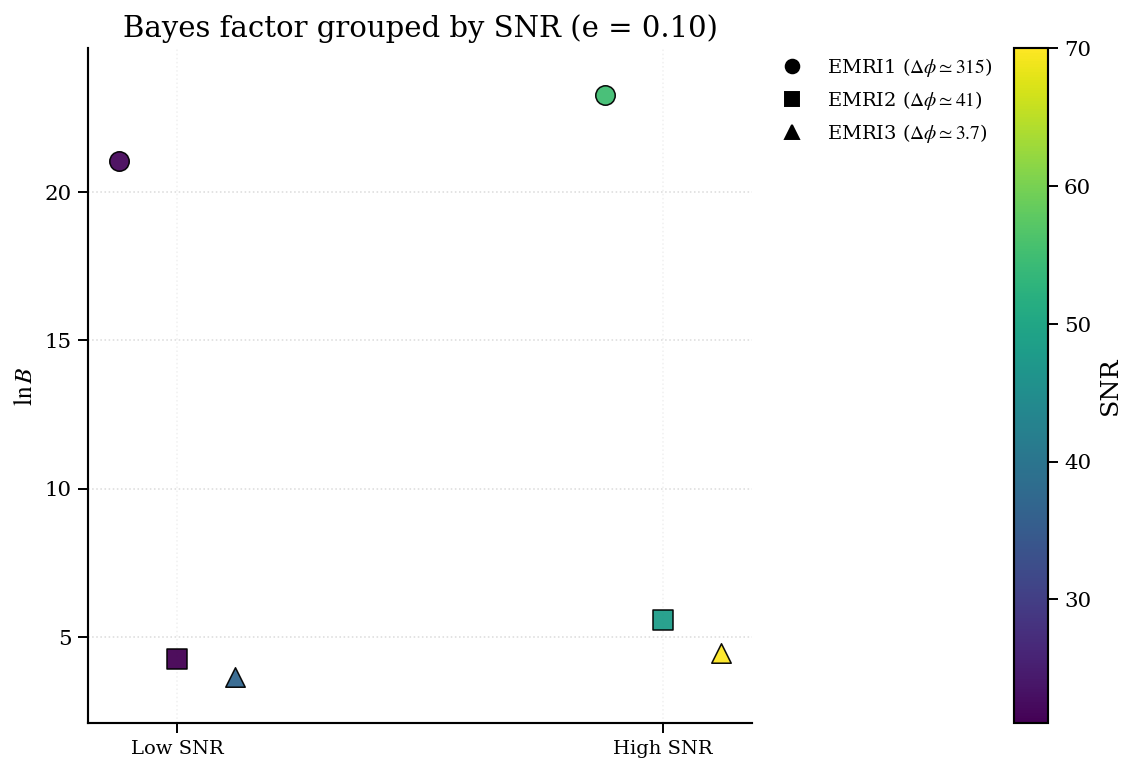}
\caption{The figure presents the logarithmic Bayes factor, $\ln B$, for different EMRI systems as a function of eccentricity in the low-SNR case. Circles, squares, and triangles denote EMRI1 ($\Delta\phi \approx 315$ radians), EMRI2 ($\Delta\phi \approx 41$ radians), and EMRI3 ($\Delta\phi \approx 3.7$ radians), respectively. The color bar indicates the corresponding SNR values.
\label{fig:3}}
\end{figure*}

Figure \ref{fig:3} shows the identification results for the presence of an accretion disk environment in three EMRI events with different eccentricities, $e=0.1$, $0.05$, and $0.01$. The results indicate that the accretion disk environment can be effectively identified in both lower  and higher eccentricity cases. Notably, for events with weaker environmental dephasing, especially EMRI2 and EMRI3, although the overall environmental dephasing is relatively small, the corresponding Bayes factor is lower in the low eccentricity case $e=0.01$, while it increases significantly as the eccentricity increases. This suggests that when the overall environmental effect is weak, residual eccentricity may improve the distinguishability of accretion disk environmental effects by introducing richer higher order harmonic structures into the waveform, thereby enhancing the model selection evidence and the corresponding Bayes factor.

\subsection{As a multimessenger source and its potential for cosmological applications}

\begin{figure*}[ht!]
\centering
\includegraphics[width=0.9\linewidth]{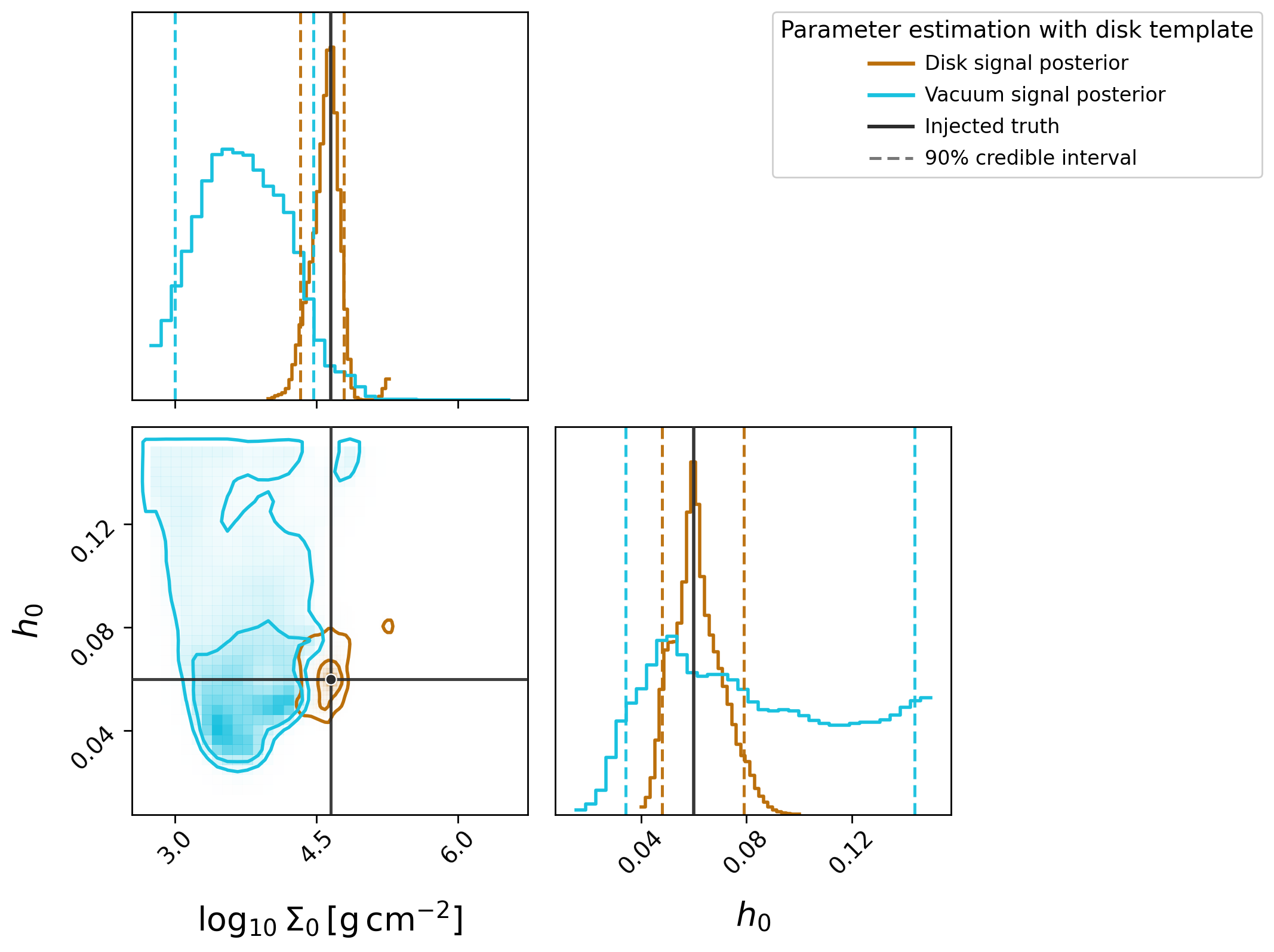}
\caption{Joint and marginalized posterior distributions of $\log_{10}\Sigma_0$ and $h_0$ for injection event EMRI3 with $e = 0.05$, comparing the disk posterior (brown) and the vacuum posterior (blue). The black solid lines indicate the injected true values, the colored solid lines denote the posterior medians, and the colored dashed lines mark the $90\%$ credible intervals ($5\%$--$95\%$).
\label{fig:5}}
\end{figure*}

Figure \ref{fig:5} presents and analyzes the constraints on the accretion disk parameters obtained with the revised model for injected vacuum and non-vacuum signals during the final four years of EMRI evolution. The results show that, for non-vacuum injections, the posterior distributions of the accretion disk parameters can be effectively constrained and exhibit a clear Gaussian-like shape, indicating that the model can identify the influence of an accretion disk in EMRIs. In contrast, for vacuum injections, the posterior distributions of the accretion disk parameters are broadly dispersed across the prior space and fail to yield meaningful constraints, demonstrating that the model does not produce spurious identifications of accretion disk effects when no such effects are present. In addition, we present in the Appendix \ref{corner} the posterior distributions of the remaining parameters obtained using the vacuum templates and the corrected templates. Table \ref{tab:2} presents the constraints on the accretion disk parameters for different EMRI events under various accretion disk parameter settings. For the three EMRI events, the injected parameters are considered $\Sigma_0 = \{5.4\times10^5,\ 5.4\times10^5,\ 4.5\times10^4\}$ g/cm$^2$ and $h_0 = \{0.015,\ 0.03,\ 0.03\}$, respectively. Combining Eqs.\ref{eq:3}--\ref{eq:h_r_wavy}, one can obtain the corresponding accretion disk parameters. The results show that even in the low SNR regime, the best constraints on the accretion disk parameters can reach better than 10\%. However, for events with weaker environmental dephasing, the constraining power becomes weaker. By contrast, if the observed event is a high SNR gold EMRI, the precision of the accretion disk parameter constraints can still be significantly improved even in the presence of weak environmental effects. Furthermore, as will be shown later, in the non-vacuum case, the dark siren method with luminosity weighting based on accretion disk parameters can further improve the precision of the Hubble constant measurement, compared with the vacuum case where only sky-localization weighting is used.

\begin{table*}[htbp]
\centering
\caption{Relative precision of the constraints on $f_{\rm Edd}$ and $\alpha$ for the three EMRI datasets.}
\label{tab:2}
\begin{tabular}{ccccccc}
\hline
Case & \multicolumn{2}{c}{EMRI1} & \multicolumn{2}{c}{EMRI2} & \multicolumn{2}{c}{EMRI3} \\
 & $f_{\rm Edd}$ & $\alpha$ & $f_{\rm Edd}$ & $\alpha$ & $f_{\rm Edd}$ & $\alpha$ \\
\hline
$e_0=0.01$, low SNR  & $2.4\%$ & $7.8\%$ & $23.3\%$ & $55.4\%$ & $79.1\%$ & $170.3\%$ \\
$e_0=0.05$, low SNR  & $2.9\%$ & $9.2\%$ & $18.9\%$ & $36.4\%$ & $52.1\%$ & $146.7\%$ \\
$e_0=0.10$, low SNR  & $5.0\%$ & $11.3\%$ & $39.4\%$ & $77.3\%$ & $63.3\%$ & $147.7\%$ \\
$e_0=0.10$, high SNR & $3.3\%$ & $7.5\%$ & $16.3\%$ & $32.6\%$ & $45.3\%$ & $74.5\%$ \\
\hline
\end{tabular}
\end{table*}

\begin{figure*}[ht!]
\centering
\includegraphics[width=0.9\linewidth]{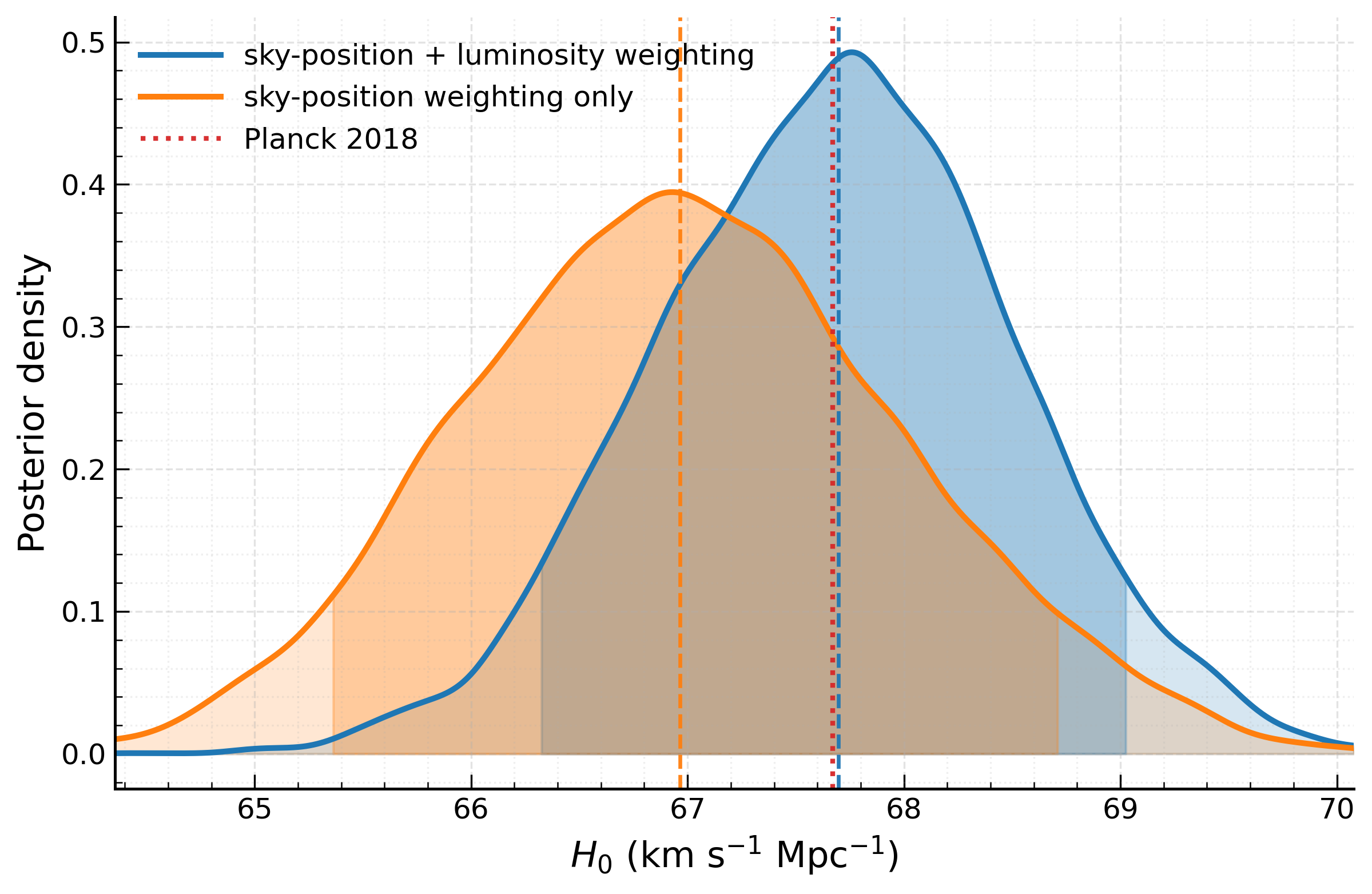}
\caption{Posterior distributions of the Hubble constant $H_0$ obtained using only sky-position weighting (orange) and by further incorporating bolometric luminosity information (blue). The colored dashed lines indicate the medians of the corresponding posterior distributions, the shaded regions denote the $90\%$ credible intervals, and the red dashed line marks the Planck 2018 reference value.
\label{fig:4}}
\end{figure*}

In this dark siren analysis, we adopt the broad-line AGN catalog published by \citep{AGNcatalog}. This catalog was uniformly constructed from galaxies and quasars with existing spectroscopic observations in SDSS DR7. It contains 14,584 AGNs within the redshift range of \(z<0.35\), covers a sky area of approximately 8200~deg\(^2\) (about 20\% of the full sky), and is highly complete within the parent spectroscopic sample of SDSS DR7. Figure \ref{fig:4} illustrates the impact on the precision of the Hubble constant constraint for wet-EMRIs with accretion disks when using sky-position weighting alone and the combination of sky-position and luminosity weighting, and compares the results obtained under these two scenarios. Compared with the result obtained using only the sky-position weighting, $H_0 = 66.97^{+1.75}_{-1.61}\,\mathrm{km\,s^{-1}\,Mpc^{-1}}$ the Hubble constant is constrained to $H_0 = 67.70^{+1.33}_{-1.37}\,\mathrm{km\,s^{-1}\,Mpc^{-1}}$, when the bolometric luminosity information is further incorporated, corresponding to an approximate \(20\%\) improvement in the constraint precision. This indicates that including bolometric luminosity information in the weighting scheme helps suppress the confusion introduced by unrelated candidate host galaxies, thereby improving the constraint on the Hubble constant.

\section{Discussion and Conclusion}\label{sec:4}
In this work, we investigate a scenario in which the small black hole in an EMRI system remains fully embedded in a radiation pressure dominated thin accretion disk during the four years before merger. We systematically incorporate into the waveform modeling the corrections to orbital evolution arising from the long term interaction between the astrophysical environment and the small black hole. On this basis, we use Bayesian parameter inference to quantitatively assess the identifiability of the EMRI environment and further explore how, once the environmental type is correctly identified, this information can improve the precision with which the Hubble constant is constrained when EMRIs are used as dark standard sirens.

Our results show that, for EMRI gravitational wave signals arising from an unknown environment, incorporating environmental effects into the waveform modeling allows one to effectively distinguish the surrounding environment, namely, to determine whether an accretion disk is present. However, for sBHs formed within the accretion disk and migrate inward, the ability to identify the environment is significantly reduced when the corresponding events have low SNR. In addition, by the time such sources enter the LISA band, their orbits are typically already highly circularized, with eccentricities often below 0.01. 
If the dephasing induced by environmental effects is also weak for example, amounting to only a few radians over a four-year observation then environmental discrimination becomes even more challenging. This behavior can be understood for two main reasons. First, previous work by \citep{PRX} has shown that for circular, equatorial EMRIs, the accretion disk parameters are often strongly degenerate. As a result, the Bayesian evidence becomes sensitive to the choice of prior range and is further reduced by a stronger Occam penalty, thereby weakening the model selection power. Second, compared with systems of higher eccentricity, low eccentricity orbits contain less information in higher order harmonics, making it more difficult to break the degeneracy among accretion disk parameters, further reducing the ability to distinguish the environmental scenario.

For the first time, we introduce the accretion disk environmental parameters measured from wet EMRIs into dark standard siren analyses as an additional constraint, demonstrating the potential of combining such information with AGN catalogs to further improve the precision of Hubble constant measurements in the future. For EMRIs, they can generally be regarded as dark sirens, since vacuum EMRIs typically lack a clear electromagnetic counterpart. As a result, in the statistical dark-siren approach, several non-host galaxies may contaminate the inference. In contrast, the accretion disk environmental information carried by wet EMRIs may provide additional guidance for selecting or weighting potential host AGNs. For example, the measured $\Sigma_0$ and $h_0$ can be combined with a specific accretion-disk model, through the relations of the disk surface density and aspect ratio, to further infer key disk parameters such as $\alpha$ and $f_{\rm Edd}$. These can then be used to estimate the plausible luminosity range of the host AGN, thereby excluding some candidate hosts whose luminosities significantly deviate from the expected range. This, in turn, can reduce the uncertainty in host association and improve the precision of the Hubble constant measurement. In this work, we further present a low eccentricity ($e=0.01$) EMRI event with moderate environmental dephasing (of order several tens of radians) as an illustrative dark siren example. Although this is only a single case study, the precision of Hubble constant measurements is expected to improve approximately as $1/\sqrt{N}$ with the number of observed events \citep{1986Na}. Therefore, only $a$ similar wet-EMRI events would be required to push the measurement precision of the Hubble constant to the sub-percent level.

On the other hand, the complex accretion disk environment surrounding non-vacuum EMRIs also suggests that they may serve as promising multimessenger sources. As the secondary inspirals inward, if the compact object is embedded within the accretion disk, its interaction with the surrounding gas may generate electromagnetic emission. 
In particular, such emission should not be limited to 
thermal emission from a circum-small black hole disk, 
but may also include (non-) thermal emission from shocks caused by the interactions between jets and the AGN disk gas, winds, and the circum-small black hole disk \citep{HT,HT2026}. 
For EMRI systems, additional electromagnetic signatures may also be triggered by gap opening by secondary and subsequent gap refilling at later stages \citep{2011_environment_effect_2}. As a result, the electromagnetic manifestation of an embedded compact object may be persistent, intermittent, or flare-like in nature; however, its detectability remains strongly dependent on the local disk environment, and direct observational evidence for such signals is still rather limited. For systems in which the orbit periodically crosses the accretion disk and triggers electromagnetic emission, such as quasi-periodic eruptions, For the QPE host systems observed so far, the central black hole masses are generally in the range of $10^5\!-\!10^6\,M_\odot$ \citep{QPEmass}. Estimates based on the observed parameters of QPE host systems indicate that the corresponding orbital radii are typically on the order of several tens to several hundreds of $r_g$ \citep{QPE11,QPE23,QPE34,QPE5,QPE6,QPE7,QPE8}. Within the framework of standard thin-disk models, the transition radius between the radiation-pressure-dominated and gas-pressure-dominated regimes, $r_{\rm gas/rad}$, depends explicitly on the accretion rate, following $r_{\rm gas/rad}\propto (\dot{m}/\epsilon)^{16/21}(M/10^{7}M_{\odot})^{2/21}$ \citep{r_from_m_dot}. Consequently, at relatively high accretion rates, the orbital radii inferred for many observed QPEs are likely to fall within the radiation-pressure-dominated regime, whereas at relatively low accretion rates, some of them may instead lie in the gas-pressure-dominated regime, or at least close to the transition between the two. Nevertheless, this conclusion still depends on the specific accretion rate and the particular disk model adopted.
At the same time, the corresponding gravitational wave frequency is often still below or only near the low frequency edge of LISA’s most sensitive band. As a result, there may still be a substantial time interval between the stage of electromagnetic outburst and the later stage when the system enters LISA’s most sensitive millihertz band. Nevertheless, if future gravitational wave observations can successfully identify an EMRI embedded in an accretion disk environment, it may still become possible to connect such an event with electromagnetic outbursts that occurred within its localized sky region over the past several decades, thereby searching for a potential electromagnetic counterpart and further extending the scientific value of EMRIs in multimessenger astronomy.

Finally, in our modeling of the orbital evolution of the sBH companion under astrophysical environmental effects, we neglected Bondi–Hoyle–Lyttleton (BHL) accretion onto the sBH within the disk \citep{BHL1}. Although this process can, to some extent, affect the orbital evolution of the sBH and its angular momentum exchange with the surrounding environment, neglecting this accretion effect is justified within the present framework, given that our analysis focuses only on the four-year evolution of the EMRI. In future studies, if one aims to characterize more accurately the impact of environmental effects on EMRI evolution and the corresponding gravitational wave signals, it will be necessary to incorporate the effects associated with accretion into a more systematic analysis.


\acknowledgments
This work was supported by the National Key R\&D
Program of China (grant No. 2021YFC2203002), the National Science and Technology Major Project (No. 2024ZD1100601), and the
National Natural Science Foundation of China (grant Nos.
12473075 and 12173071). This work made use of the High-
Performance Computing Resource in the Core Facility for
Advanced Research Computing at Shanghai Astronomical
Observatory.

\bibliography{references}

\begin{figure*}[ht!]
\centering
\includegraphics[width=0.9\linewidth]{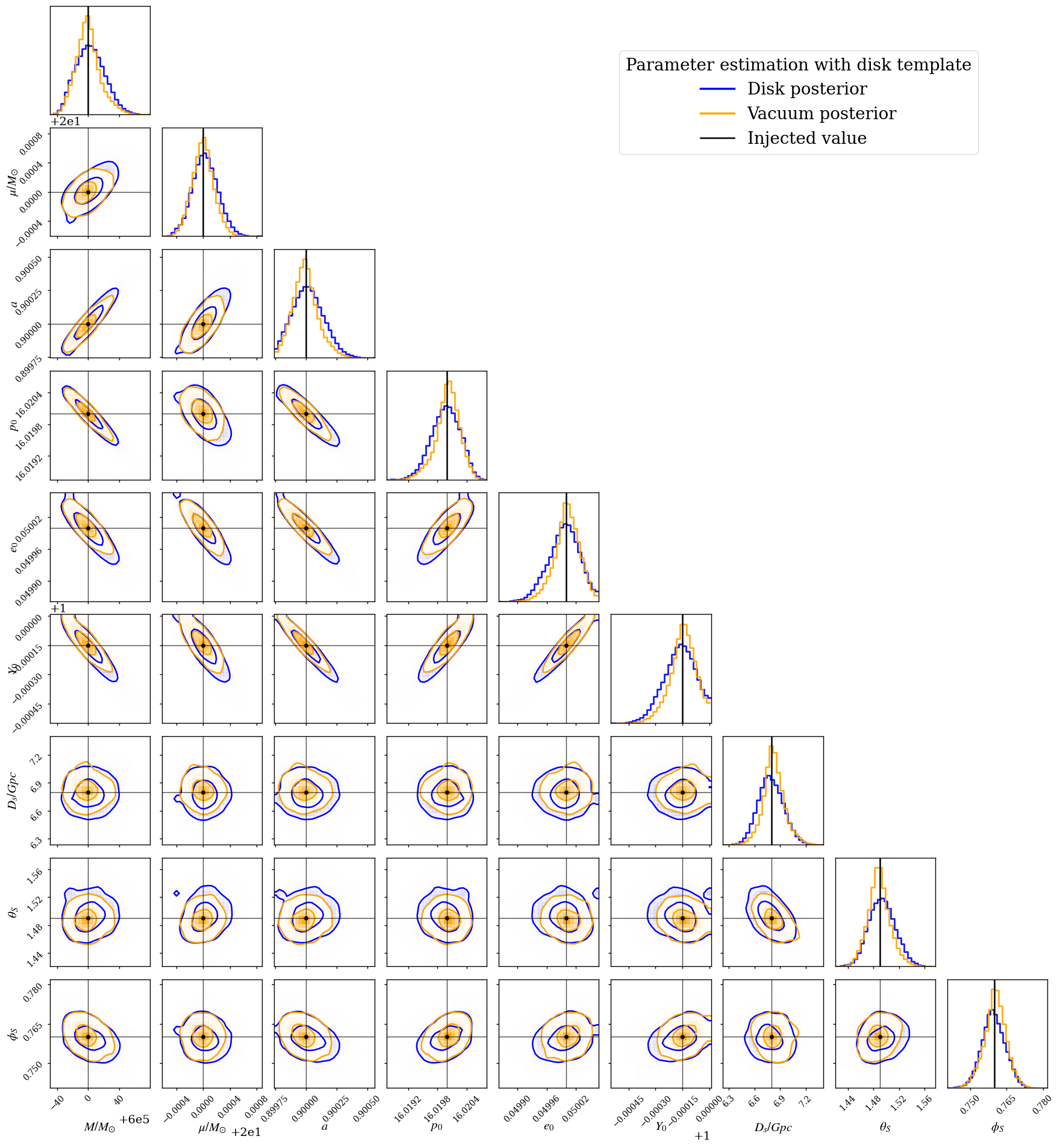}
\caption{Corner plots showing the posterior probability distributions of the recovered parameters obtained with the disk template for both the disk injected and vacuum injected signals, based on a four year LISA observation. The injected source parameters are $(M=5\times10^5,\ \mu=20,\ a=0.9,\ p=16.02,\ e=0.05,\ \cos\iota=0.99985,\ d_L=6.8\,\mathrm{Gpc},\ \theta_S=1.49,\ \phi_S=0.76)$. For the disk-injected signal, the accretion disk parameters are further set to $(\Sigma_0=4.5\times10^54,\ h_0=0.03)$. The blue curves denote the posterior distributions for the disk-injected signal, whereas the yellow curves represent those for the vacuum-injected signal. The black lines indicate the injected values, and the different contours correspond to the 68\% and 90\% confidence regions.
\label{fig:disk_vorner}}
\end{figure*}

\begin{figure*}[ht!]
\centering
\includegraphics[width=0.9\linewidth]{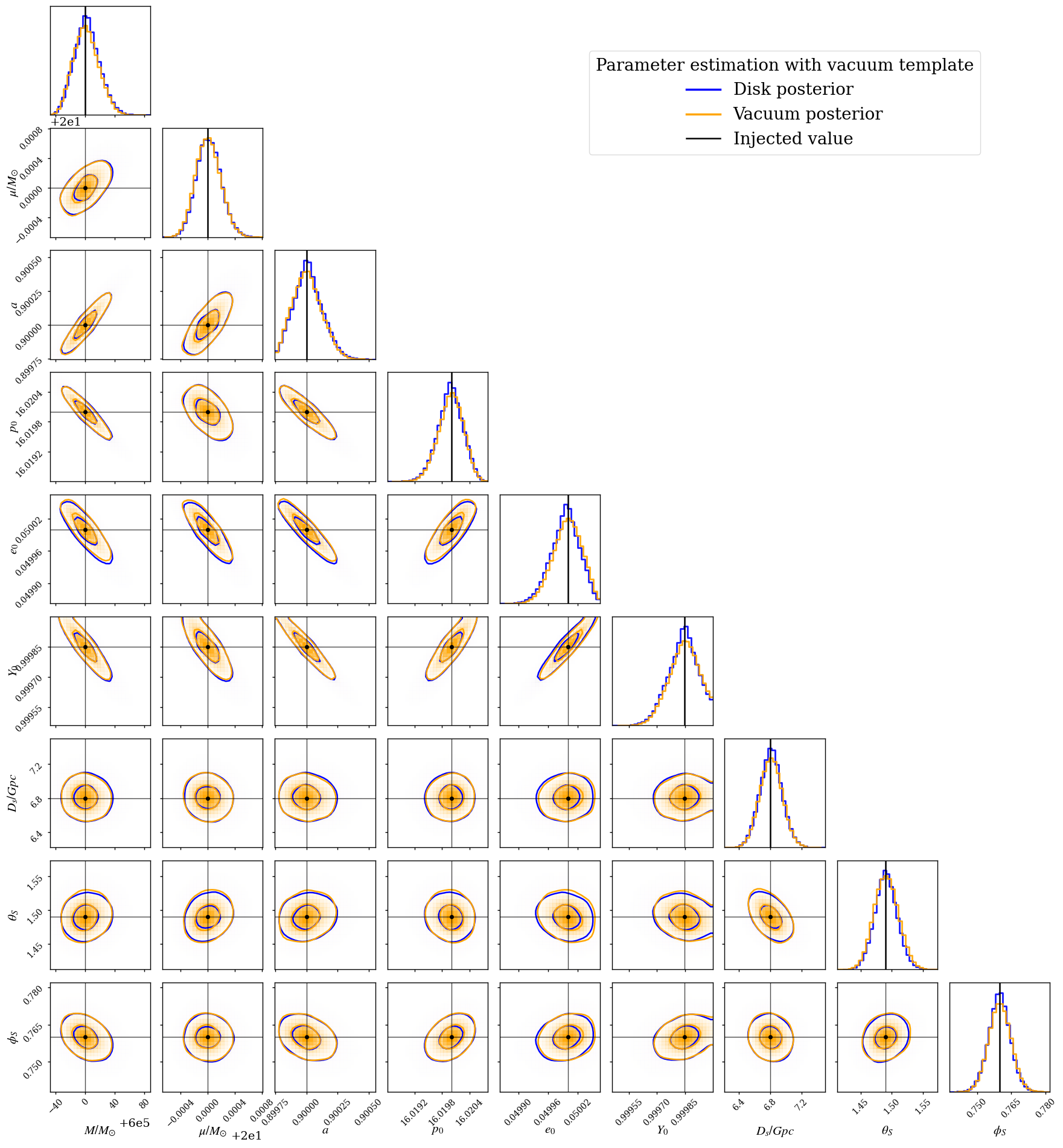}
\caption{Corner plots showing the posterior probability distributions of the recovered parameters obtained with the vacuum template for both the disk injected and vacuum injected signals, based on a four year LISA observation. The injected source parameters are $(M=5\times10^5,\ \mu=20,\ a=0.9,\ p=16.02,\ e=0.05,\ \cos\iota=0.99985,\ d_L=6.8\,\mathrm{Gpc},\ \theta_S=1.49,\ \phi_S=0.76)$. For the disk-injected signal, the accretion disk parameters are further set to $(\Sigma_0=4.5\times10^54,\ h_0=0.03)$. The blue curves denote the posterior distributions for the disk-injected signal, whereas the yellow curves represent those for the vacuum-injected signal. The black lines indicate the injected values, and the different contours correspond to the 68\% and 90\% confidence regions.
\label{fig:vacuum_corner}}
\end{figure*}

\appendix\label{corner}

\end{document}